# Josephson vortex lattice melting in Bi-2212 probed by commensurate oscillations of Josephson flux-flow

Yu.I. Latyshev, V.N. Pavlenko, A.P. Orlov
*Institute of Radio-Engineering and Electronics, Russian Academy of Sciences, Mokhovaya 11-7, 101999 Moscow, Russia*
X. Hu
*National Institute for Materials Science, 1-2-1 Sengen – Tsukuba, 305-0047 Ibaraki, Japan*

**Abstract.** We studied the commensurate semi-fluxon oscillations of Josephson flux-flow (JFF) in a Bi-2212 stacked structures near $T_c$ as a probe of melting of Josephson vortex lattice (JVL). We found that oscillations exist above 0.5T. The amplitude of oscillations is found to decrease gradually with temperature and to turn to zero without any jump at $T$=$T_0$, 3.5K below the resistive transition temperature $T_c$ indicating the phase transition of the second order. This characteristic temperature $T_0$ is identified as the Berezinskii-Kosterlitz-Thouless (BKT) transition temperature, $T_{\mathrm{BKT}}$, in elementary superconducting layers of Bi-2212 at zero magnetic field. On the base of these facts we infer that melting of triangular JVL occurs via the BKT phase with formation of characteristic flux loops containing pancake vortices and anti-vortices. The B-T phase diagram of the BKT phase found out from our experiment is consistent with theoretical predictions.

**Introduction.** The vortex phase diagram in layered high-$T_c$ materials in parallel magnetic fields is significantly less studied than for perpendicular fields. That is related with a great difficulty in visualization of Josephson vortices [1] and JVL [2]. Recently the method of identification of triangular JVL has been found [3]. That was demonstrated as oscillations of Josephson flux-flow resistance in narrow Bi-2212 mesa structures in parallel field with periodicity of $\frac{1}{2}\Phi_0$ per elementary Josephson junction. The effect has been interpreted as a result of commensurability of triangular JVL period with mesa width. In this paper (see also [4]) we develop an idea to use the effect of JFF commensurate oscillations for probing of JVL melting. In a triangular lattice the periodic rows of JVs in adjacent layers are shifted by *p*. Therefore JFF oscillations with semi-fluxon periodicity reflect the transverse coherence of triangular lattice. The melting of triangular lattice should be accompanied by a loss of transverse coherence and, as a result, by disappearance of semi-fluxon oscillations. Different scenarios of melting can happen, melting into liquid phase or melting into the BKT phase [5-7]. For the latter case the possibility of second order transition has been considered [7,8]. The influence of BKT transition on JVL phase at high fields and high temperatures has been widely debated [5-9]. However, until recently there were no systematic experimental studies of JVL melting.

**Results and discussion.** The experiment was carried out on the stacked structures of slightly overdoped Bi-2212 with lateral sizes $L_a$ x $L_b$ =15-30μm x 3-5μm and containing about 100 elementary junctions. The structures have been fabricated by double sided processing of Bi-2212 whiskers by focused ion beam (FIB) [10]. The field was oriented strictly parallel to the *ab*-plane and perpendicular to the *a*-axis. The parallelity of the field orientation was adjusted within 0.01° by fixing that at the sharp maximum of JFF magnetoresistance with field rotation around the *a*-axis. For that purpose along with a main coil providing magnetic field up to 1T additional perpendicular coil has been used. The data have been collected by computer controlled current source and nanovoltmeter.

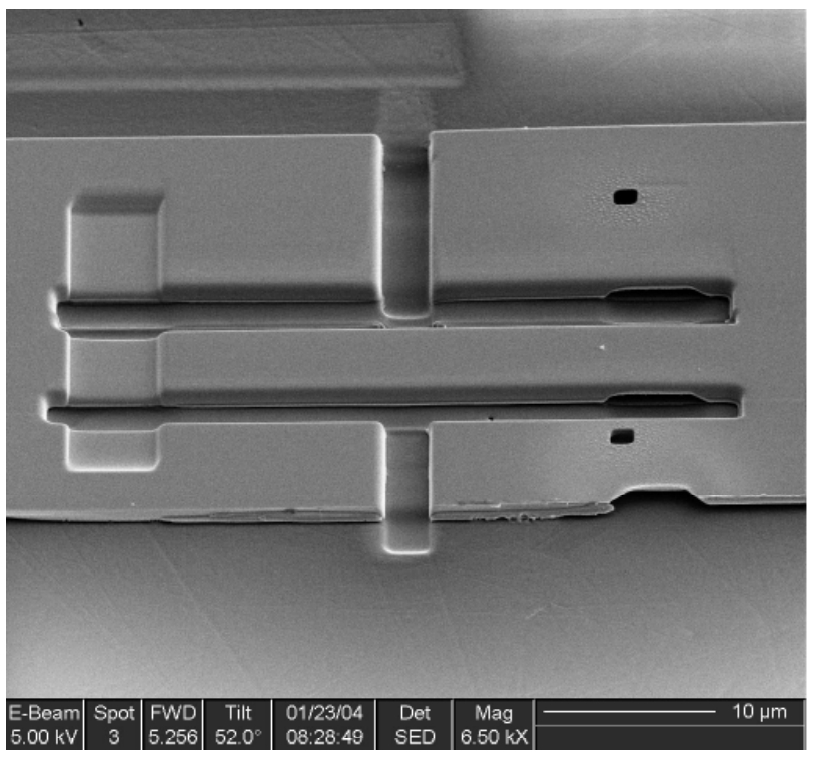

Fig. 1. SEM picture of the stacked structure fabricated by double sided FIB etching of Bi-2212 single crystal whisker.

We traced oscillating dependence of JFF resistance on parallel magnetic field at fixed temperatures with temperature variation by small steps (Fig. 2). The period of oscillations exactly corresponds to ½ fluxon per elementary junction, $\Delta B$=0.5$\Phi_0$/ $L$ $s$, with $L$ the stack size perependicular to magnetic field and $s$ the interlayer spacing. The amplitude of oscillations decreases with temperature and turns to zero at some temperature $T_0$, 3.5K below $T_c$ (Fig. 3). Fig.2 shows that at the fixed temperature oscillations exist within some field interval. The boundaries of that interval, marked at the picture, define lower and upper boundaries of triangular JVL state at the B-T diagram (Fig.4).

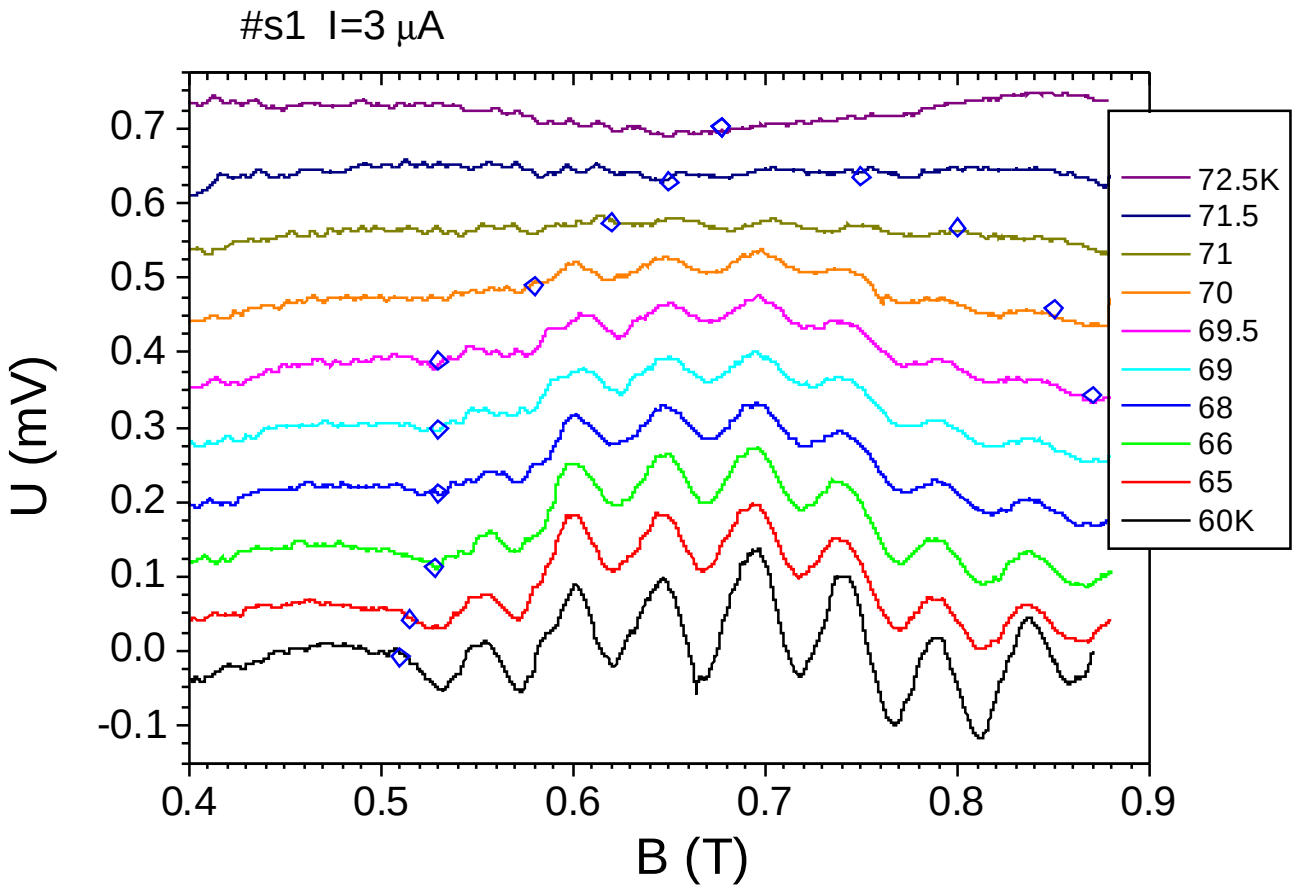

Fig. 2. Semifluxon oscillations of Josephson flux-flow voltage of Bi-2212 mesa #s1with lateral sizes $L_a$ x $L_b$ = 15 µm x 5 µm in parallel field B//b. The linear part is substracted. The curves are shifted for clarity. The markers (rhombuses) indicate the lower and upper boundaries of oscillations existence.

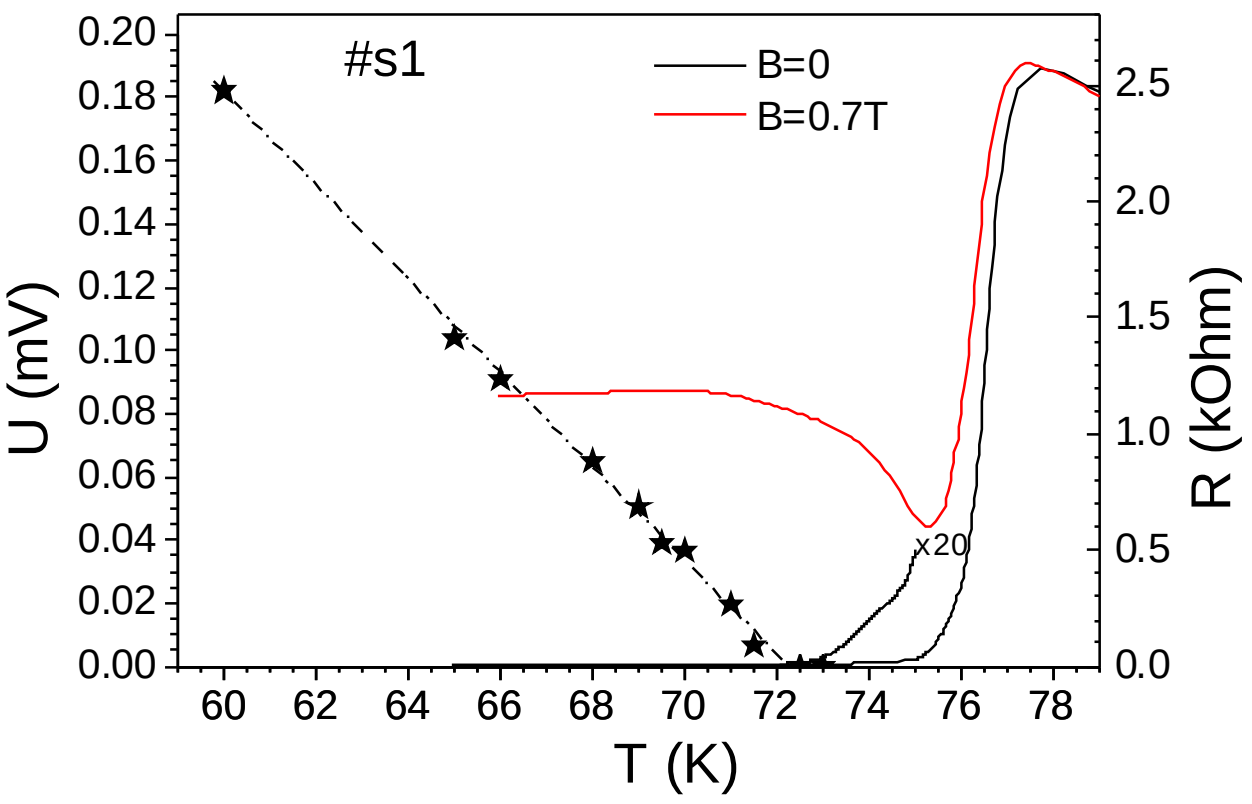

Fig. 3. Temperature dependence of the amplitude of semifluxon JFF oscillations on parallel magnetic field measured near the field 0.7T for Bi-2212 stacked junction # s1 and superconducting transition for the same junction at zero field and at B=0.7T.

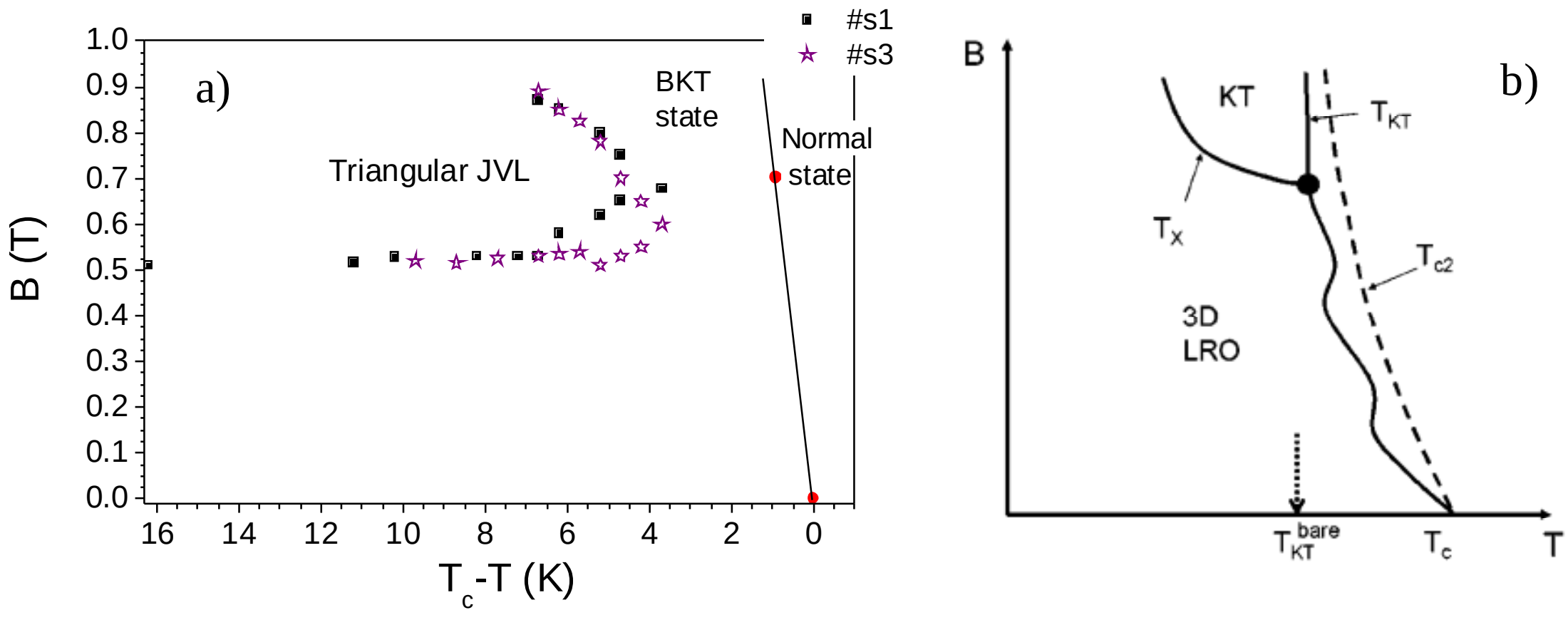

Fig. 4. Phase diagram of the JVL state restored from the data on semifluxon oscillations of JFF voltage on parallel magnetic field for two Bi-2212 mesas (a) and schematic picture of phase diagram considered in [7] (b).

Let discuss the main features observed. The characteristic point at the phase diagram, corresponds to *B*= 0.6-0.7T, where upper and lower boundaries meet each other at $T=T_0$. There are no JFF oscillations above this temperature. We can then conclude that there is no triangular JVL state above that temperature. The value of $T_0$ lies 3.5 K below the transition temperature. That is very close to the bare BKT transition temperature observed at zero magnetic field on similar Bi-2212 single crystals [11]. The BKT transition is characterized by spontaneous formation of the free pairs of pancake vortices and anti-vortices within elementary superconducting layers. Below $T_{BKT}$ the vortex – antivortex pairs can be unbound by the in-plane current and the I-V characteristics have the power law $V=I^{a(T)+1}$ where exponent $a = \Phi_0^2 d/(4\pi\lambda)^2$, with *d* the thickness of the layer and $\lambda$ the London penetration depth, is proportional to the unbinding energy. At the BKT transition this exponent undergoes a universal jump from 2 to 0, known as the Nelson-Kosterlitz jump, that is a characteristic feature of the BKT transition. By observation of this jump the BKT transition has been identified in the elementary conducting layers of Bi-2212 single crystals [12] with $T_{BKT}-T_{co}$=3.5K.

The BKT transition in parallel magnetic field [5, 9] was considered in connection with melting of JVL [7,8] and with observation of Lorentz force independent dissipation when both transport current and magnetic field lie in the *ab*–plane [13]. The elementary process accompanying JVL melting has been considered by Blatter et al. [5]. That is the hopping of a segment of JV into the neighbour junction with formation of a loop that includes pancake vortex and anti-vortex (Fig.5).

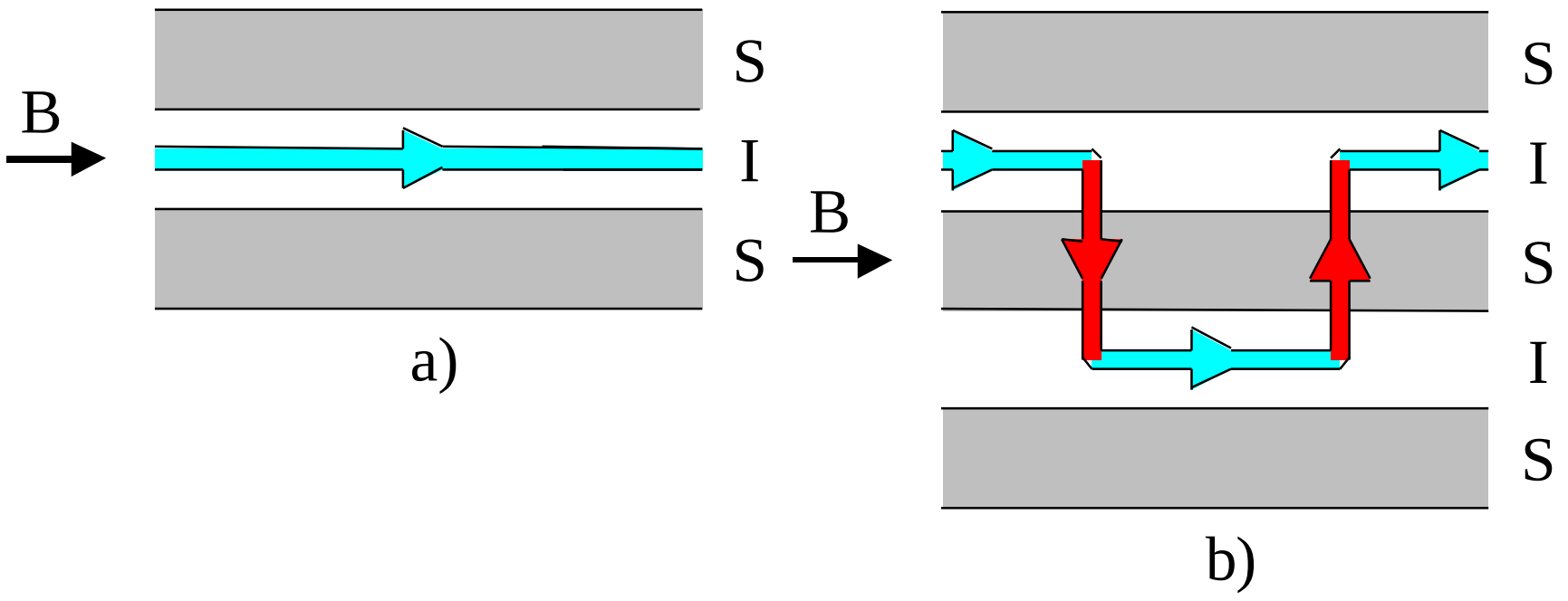

Fig. 5. The schematic illustration of hopping of a Josephson vortex segment with formation of a flux loop containing a pair of pancake vortex and anti-vortex (marked by red colour): (a) Josephson vortex before hopping, (b) after hopping. Josephson vortex marked by blue colour.

The BKT transition facilitates the hopping providing free vortices and anti-vortices necessary for formation of a loop. The unbinding of vortex-antivortex pairs by in-plane component of the current circulating around Josephson vortex happens even at temperatures considerably lower than $T_{BKT}$. The hopping becomes preferable at higher temperature because of an increase of thermal fluctuations and also at high enough magnetic field with increase of vortex concentration and, correspondingly, with an increase of the inter-vortex repulsive interaction in one junction. The critical field for JVL melting into the BKT phase *B** has been calculated by Hu and Tachiki [7,8] $B^* = \Phi_0 /(2\sqrt{3}g\, s^2)$ with $g$ the anisotropy of London penetration depths $g=l_c/l_{ab}$, $s$ the spacing between elementary conducting layers. They also considered B-T phase diagram of the JVL melting schematically shown in Fig. 4b.

Experimentally found phase diagram of JVL melting is quite similar to the theoretical picture. The maximum temperature of existing the BKT phase, $T_0$ as it was mentioned well corresponds to the bare BKT transition. The critical magnetic field *B** estimated for our samples with $g$=500 [14] corresponds to 0.5*T* which is close to the experimental value 0.6-0.7*T*. The upper boundary of JVL existence is also in qualitative agreement with theoretical *B*(*T*) dependence for melting line: *B* grows with *T* decrease. The crossing of the BKT melting line by moving from JVL state either by increasing temperature or field corresponds to continuous decrease of the amplitude of oscillations to zero without any jump expected for the first order transition. That is a signature of the JVL melting into the BKT phase since, as it was argued in [7] for $g > 9$ and $B>B^*$, that should be a phase transition of the second order. Note that experimental picture corresponds to the sliding JVL, while theoretical picture relates to the static case. However, due to the small dc currents used in experiment, JVL velocity was relatively small, about 3% of Swihart velocity.

To the contrast the lower boundary is characterized by much sharper variation of the oscillation amplitude with a decrease of field. The origin of low boundary is still not quite clear. The JFF voltage is known to have a threshold as a function of magnetic field [15]. That corresponds to flux density 0.7 $\Phi_0$ per junction. The oscillations appear starting with field of about 0.5 T nearly independent on temperature. That field corresponds to 5-7 periods of triangular JVL. That appears to be the threshold value for commensurability effect corresponding to the minimal number of periods for the lattice starts to behave as a piece of solid.

We found that the Josephson flux-flow branch still exists on the I-V characteristic above $T_{BKT}$, however, JFF voltage drops down rapidly at $T > T_{BKT}$ (Fig.2). The study of JFF state above $T_{BKT}$ is of great interest for the future research. The interesting point is also to study the influence of *c*-axis field component on JVL melting temperature.

**Acknowledgement.** We acknowledge the fruitful discussions with L.N. Bulaevskii, V. Geshkenbein and M.B. Gaifullin. The work has been supported by CRDF grant No. RP1-2397-MO-02, the Russian Ministry of Science and Industry grant No. 40.012.1.111.46, Russian state program for development of new materials and RFBR-CNRS grant No. 03-02-22001.